\documentclass[journal]{IEEEtran}
\IEEEoverridecommandlockouts
\usepackage{amsmath,amssymb,amsfonts}
\usepackage{algorithmic}
\usepackage{graphicx}

\ifCLASSOPTIONcompsoc \usepackage[caption=false,font=normalsize,labelfont=sf,textfont=sf]{subfig}
\else
\usepackage[caption=false,font=footnotesize]{subfig}
\fi

\usepackage{accents}
\usepackage{textcomp}
\def\BibTeX{{\rm B\kern-.05em{\sc i\kern-.025em b}\kern-.08em
    T\kern-.1667em\lower.7ex\hbox{E}\kern-.125emX}}

\usepackage{lipsum} 
 \usepackage{balance}
\usepackage{booktabs} 
\usepackage{enumitem}
\usepackage{mathtools}
\usepackage[lined,boxed,commentsnumbered,linesnumbered,ruled]{algorithm2e}
\usepackage{tikz} 
\usetikzlibrary{shapes, patterns,decorations.text, decorations.pathreplacing}
\usetikzlibrary{external}
\usepackage{soul}

\usepackage[noadjust]{cite} 

\usepackage{pgfplots}
\pgfplotsset{filter discard warning=false}
\pgfplotsset{compat=1.14}
\usepgfplotslibrary{colorbrewer}
\usepgfplotslibrary{groupplots}

\usepackage[capitalize]{cleveref}
\crefname{equation}{\unskip}{\unskip}
\crefname{claim}{Claim}{Claims} 
\usepackage{array}
\usepackage{multirow}
\newcolumntype{C}[1]{>{\centering\arraybackslash}p{#1}}
\allowdisplaybreaks

\newcommand{\E}[1]{\mathbb{E}\left[ #1 \right]}

\newcommand{\Var}[1]{\operatorname{Var} \left[ #1 \right]}

\newcommand{\rmean}{\bar{r}_1(\tau)}
\newcommand{\rprocess}{r_1(\tau)}

\newcommand{\estar}{\epsilon^*}
\newcommand{\sstart}{\alpha}
\newcommand{\ssend}{\beta}
\newcommand{\ssrange}{\left[\sstart, \ssend\right]}

\newcommand{\e}{\epsilon}

\newcommand{\m}{\mu_0}
\newcommand{\mb}{\breve{\mu}_0}
\newcommand{\defn}{\triangleq}
\newcommand{\der}{\mathrm{d}}
\newcommand{\ti}{\to\infty}
\newcommand{\grows}{\ti}

\newcommand{\FER}{P_\mathsf{f}}
\newcommand{\FERt}{P_\mathsf{f,t}}
\newcommand{\FERu}{P_\mathsf{f,u}}
\newcommand{\BER}{P_\mathsf{b}}
\newcommand{\BERt}{P_\mathsf{b,t}}
\newcommand{\BERu}{P_\mathsf{b,u}}
\newcommand{\BLER}{P_\mathsf{bl}}
\newcommand{\BLERt}{P_\mathsf{bl,t}}
\newcommand{\BLERu}{P_\mathsf{bl,u}}
\newcommand{\doublehat}[1]{\tilde{#1}}

\newcommand{\gammasingle}{\breve{\gamma}}
\newcommand{\gammadouble}{\doublehat{\gamma}}
\newcommand{\sstartsingle}{\breve{\sstart}}
\newcommand{\sstartdouble}{\doublehat{\sstart}}

\newcommand{\ssenddouble}{\doublehat{\ssend}}
\newcommand{\thetasingle}{\breve{\theta}}
\newcommand{\thetadouble}{\doublehat{\theta}}
\newcommand{\nusingle}{\breve{\nu}}
\newcommand{\nudouble}{\doublehat{\nu}}

\newcommand{\dv}{d_\mathsf{v}}
\newcommand{\dc}{d_\mathsf{c}}

\newcommand{\fsingle}{f^{\mathsf{(1)}}_{\tau_0}}
\newcommand{\fdouble}{f^{\mathsf{(2)}}_{\tau_0}}

\newcommand{\sstartlb}{\sstartdouble_\mathsf{LB}}

\newcommand{\Pfone}{\FER^{\mathsf{ph}_1}}
\newcommand{\Pbone}{\BER^{\mathsf{ph}_1}}
\newcommand{\Pblone}{\BLER^{\mathsf{ph}_1}}
\newcommand{\Ptftwo}{\accentset{\circ}{P}_{\mathsf f}^{\mathsf{ph}_2}}
\newcommand{\Pbtwo}{\BER^{\mathsf{ph}_2}}
\newcommand{\Pbltwo}{\BLER^{\mathsf{ph}_2}}

\newcommand{\speed}{s}
\newcommand{\speede}{\speed_\e}

\providecommand{\keywords}[1]
{
    {\small	\textbf{\textit{Index Terms---}#1.}}
}

\begin{document}

\title{Finite-Length Scaling of Spatially Coupled LDPC Codes Under Window Decoding Over the BEC}
\author{
    Roman Sokolovskii, \IEEEmembership{Graduate Student Member, IEEE}, Alexandre Graell i Amat, \IEEEmembership{Senior Member, IEEE}, \\and Fredrik Br\"annstr\"om, \IEEEmembership{Member, IEEE}
    \thanks{This paper was presented in part at the IEEE Information Theory Workshop~(ITW), Visby, Gotland, Sweden, August 2019.}
    \thanks{This work was funded by the Swedish Research Council (grant 2016-4026).}
    \thanks{R. Sokolovskii, A. Graell i Amat, and F. Br\"annstr\"om are with the Department of
    Electrical Engineering, Chalmers University of Technology, SE-41296 Gothenburg, Sweden (email:
    \{roman.sokolovskii,alexandre.graell,fredrik.brannstrom\}@chalmers.se).}}

\maketitle

\begin{abstract}
    We analyze the finite-length performance of spatially coupled low-density
    parity-check \mbox{(SC-LDPC)} codes under window decoding over the binary erasure channel.
    In particular, we propose a refinement of the scaling law by Olmos and Urbanke for the frame error rate (FER) of terminated
    SC-LDPC ensembles under full belief propagation (BP) decoding.
    The refined scaling law models the decoding process as two independent Ornstein-Uhlenbeck
    processes, in correspondence to the two decoding waves that propagate toward the center of the coupled chain for
    terminated SC-LDPC codes.
    We then extend the proposed scaling law to predict the performance of (terminated) SC-LDPC code ensembles under the
    more practical sliding window decoding.
    Finally, we extend this framework to predict the bit error rate (BER) and block error rate (BLER) of SC-LDPC code ensembles.
     The proposed scaling law yields very accurate predictions of the FER, BLER, and BER for both full BP and window decoding.
\end{abstract}

\section{Introduction}
Spatially coupled low-density parity-check (SC-LDPC) \mbox{codes~\cite{Jime99,Lent10}} are remarkable for two reasons: First, they exhibit
\textit{threshold saturation}---suboptimal belief propagation (BP) decoding of
an SC-LDPC code can achieve the decoding threshold of optimal maximum a posteriori (MAP) decoding of the
underlying uncoupled ensemble.
The threshold saturation effect, first observed in~\cite{Lent10}, was proved for the binary erasure
channel (BEC) in~\cite{Kude11} and for the more general class of binary-input memoryless symmetric
channels in~\cite{Kude13}.
Second, spatial coupling preserves the distance growth properties of the underlying uncoupled ensemble.
Thus, the minimum distance of a regular SC-LDPC ensemble grows linearly with the block length~\cite{Srid07}.
In other words, spatial coupling allows for both improved iterative decoding thresholds and low error
floors~\cite{Cost14}.
The concept of spatial coupling extends beyond the realm of low-density parity-check (LDPC) codes; it has been
successfully applied in the context of, e.g., turbo-like codes~\cite{Molo17} and product-like codes \cite{Smit12}, as
well as to lossy compression~\cite{Are12} and compressed sensing~\cite{Don13}.

Spatial coupling consists of interconnecting a sequence of Tanner graphs of the underlying uncoupled
codes according to a predefined pattern.
The key to improved asymptotic performance of SC-LDPC codes is the
\textit{structured irregularity} at the boundaries of the resulting coupled chain due to termination: the lower
average degrees of the check nodes (CNs) at the boundaries of the chain result in the presence of stronger
subcodes, from which reliable information propagates during BP decoding toward the center
of the chain in a wave-like fashion.
The termination is associated with some rate loss, which tends to zero as the chain length grows
large. To fully exploit threshold saturation and limit the rate loss, SC-LDPC codes require a large chain length. This, however, results in an unacceptably high decoding delay under full BP decoding (i.e., when BP decoding is applied to the whole chain). To limit the decoding delay,  
so-called window decoding, where decoding is limited to a window of few spatial positions that slides over the chain, is used in practice. Window decoding of SC-LDPC codes was originally proposed in~\cite{Iye12}. 

The analytical prediction of the error probability of SC-LDPC codes for a given finite code length is a research
problem of practical interest.
To that end, Amraoui \textit{et al.}~\cite{Amra09} proposed a finite-length scaling law for
uncoupled LDPC code ensembles over the BEC that accurately predicts the frame error rate (FER) in
the waterfall region.
The scaling law is based on the analysis of the sequence of residual
graphs obtained during peeling decoding (equivalent to BP decoding for the BEC).
Some extensions of the scaling law to more general channels were presented
in~\cite{Ezri08_slope,Ezri08_shift}.
Following a similar approach, a scaling law for terminated SC-LDPC ensembles was proposed
in~\cite{Olmo15}.
The authors modeled the stochastic process associated with the fraction of degree-one CNs during
peeling decoding by an appropriately chosen Ornstein-Uhlenbeck process.
The probability of decoding error is then predicted using the probability distribution of the earliest
time when peeling decoding runs out of degree-one CNs, which, in turn, is obtained from an
approximation of the first hit time of the Ornstein-Uhlenbeck process.
The parameters of this process are estimated from a system of coupled differential equations dubbed
\textit{mean and covariance evolution}.
The framework proposed in~\cite{Olmo15} was applied to the case of spatially coupled
protograph-based LDPC code ensembles in~\cite{Stin16} and suggested for generalized spatially coupled
LDPC ensembles in~\cite{Cost18}.
Unfortunately, unlike in the case of uncoupled LDPC ensembles in~\cite{Amra09}, the FER predictions
in~\cite{Olmo15} show a relatively significant mismatch with respect to the simulated
curves.
This mismatch was explained in~\cite{Olmo15} by the inadequacy of the used exponential approximation
of the first hit time distribution of the Ornstein-Uhlenbeck process. Furthermore, an important limitation of the works \cite{Olmo15,Stin16,Cost18} is that only full BP decoding is considered, while in practice SC-LDPC codes are decoded using a sliding window decoder.

In this paper, we propose a scaling law to predict the finite-length performance of (terminated) SC-LDPC ensembles under
window decoding over the BEC.
In particular, for full BP decoding, we propose a refinement of the scaling law for the FER of terminated SC-LDPC codes
proposed in~\cite{Olmo15} that results in a much better FER prediction, closing the gap between analytical and simulated curves.
The proposed refined scaling law is based on modeling the decoding process as two
independent Ornstein-Uhlenbeck processes that correspond to the two decoding waves propagating
toward the center of the coupled chain from the termination boundaries, as opposed to the scaling law in~\cite{Olmo15}, which assumes a single process. Accordingly, we model the probability density function (PDF) of the first hit time of the resulting
process as the convolution of two exponential PDFs, yielding the PDF of an Erlang distribution,
which is used to predict the probability of decoding error.
We further improve the match between the predicted performance and simulation results by introducing
a dependency on the channel parameter of the underlying scaling constants that can be computed from the mean
evolution.
We also adapt the scaling law to predict the bit error rate (BER) and block error rate (BLER) performance of
SC-LDPC ensembles.
Finally, we extend the scaling law to window decoding of (terminated) SC-LDPC ensembles.
The key idea is to observe that in this case decoding unfolds in two stages: in the
first, there is only one decoding wave; in the second, there might be two waves.
The proposed framework allows for an accurate prediction of the error rate of finite-length
SC-LDPC code ensembles under window decoding.

\section{Preliminaries}
\label{sec:preliminaries}

\begin{figure}
    \centering
    \includegraphics[width=\columnwidth]{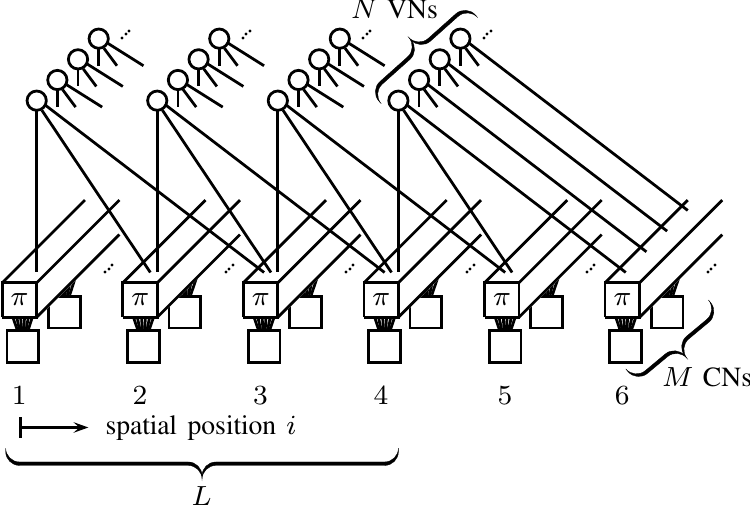}
    \vspace{-15pt}
    \caption{Tanner graph of the terminated $(\dv,\dc,L,M)$ SC-LDPC ensemble with $\dv=3,\dc=6,$
    $L=4$, and $M$ CNs and $N$ VNs per spatial position.}
    \label{fig:sc_ensemble}
    \vspace{-2ex}
\end{figure}

We consider the $(\dv,\dc,L,N)$ SC-LDPC code ensemble introduced in~\cite{Olmo15}, whose Tanner
graph is depicted in Fig.~\ref{fig:sc_ensemble}.
The Tanner graph is constructed by placing $L$ copies of a $(\dv,\dc)$-regular LDPC code of variable
node (VN) degree $\dv$ and CN degree $\dc$ in $L$ spatial positions in the set
$\mathcal{L}=\{1,\ldots,L\}$.
Each spatial position consists of $N$ VNs and $M\!=\!\frac{\dv}{\dc}N$ CNs, where we assume $M$
is an integer.
We denote by $L$ the coupling length and by $N$ the component code length.
The set of all $LN$ VNs in the Tanner graph, i.e, the set of all code bits, is referred to as the \textit{frame,} and the set of $N$ VNs at a spatial position as a \textit{block}.
The $L$ copies are then coupled as follows: each VN at position $i\in \mathcal L$ is connected to
one CN chosen uniformly at random at each of the positions in the range $[i,\ldots,i+\dv-1]$.
To connect the overhanging edges at the end of the chain, $\dv-1$ additional positions containing
CNs only are appended, resulting in a \textit{terminated} ensemble.
The detailed procedure of generating elements from this ensemble is described in~\cite{Olmo15}.
Note that the ensemble is structured from the VN perspective: each VN is connected to
$\dv$ different spatial positions.
In contrast, no structure is enforced on the connectivity of the CNs---a CN at position $i\in\{\dv,\ldots,L\}$ can be
connected to $\dc$ VNs from an arbitrary non-empty subset of positions in the range $[i-\dv+1,\ldots,i-1,i]$.
This particular ``semi-structured'' connectivity was considered in~\cite{Olmo15} in place of the more
conventional ensemble with \emph{smoothing} parameter~\cite{Kude11} to simplify the analysis.
Besides the terminated ensemble, we also consider the following two ensembles: the \emph{truncated} ensemble, where no additional positions containing CNs only are added, resulting in VNs at the end of the coupled chain with lower degree; and the \emph{unterminated} ensemble, where the ensemble is neither terminated nor truncated, resulting in a virtually infinite sequence of coupled codes. For the latter, we may consider the evaluation of the error probability over the first $L'$ positions of the coupled chain.

The excellent performance of SC-LDPC codes stems from the lower degree of the CNs at the boundaries of the coupled chain; the left boundary in the case of the truncated and unterminated ensembles, and both boundaries in the case of a terminated ensemble. In particular, the BP decoding of truncated and unterminated SC-LDPC codes under full BP decoding (i.e., the decoding is performed block-wise over the whole Tanner graph) is characterized by a wave-like decoding effect where a decoding wave propagates from the left boundary of the coupled chain rightwards. In the case of a terminated SC-LDPC code and full BP decoding, two waves propagate from the exterior of the coupled chain to the interior. 

To alleviate the inherent large decoding latency of SC-LDPC codes under full BP decoding, a window decoder \cite{Iye12} that exploits their convolutional structure is typically used in practice. The window decoder restricts decoding to CNs in a window of $W$ spatial positions and has a decoding latency of $N(W+\dv-1)$ bits. After a prescribed number of decoding iterations, a decision on the bits in the left-most spatial position is made and the window \emph{slides} one position to the right over the Tanner graph. As the window size grows, performance tends to that of full BP. Note that under window decoding, the decoding of terminated SC-LDPC codes is also characterized by a single decoding wave that propagates with the sliding window, except when the window hits the end of the chain, in which case two waves propagate within the window. This two-phase phenomenon will be exploited when analyzing the finite-length scaling of SC-LDPC codes under window decoding.

For the analysis, we will consider decoding using the peeling decoding algorithm~\cite{Luby97}.
On the BEC, the peeling decoder is equivalent to the BP decoder, in the sense that it is bewildered
by the same stopping sets and hence (for an infinite number of iterations) yields identical
performance.
However, peeling decoding makes the finite-length scaling analysis more tractable.
The initialization step of peeling decoding consists of removing all non-erased VNs and adjacent
edges from the Tanner graph.
At every subsequent iteration, one degree-one CN is randomly selected.
Since the connected VN is known and the code bit can be recovered, the chosen CN is removed from the
graph along with the neighbor VN and all $\dv$ connected edges.
Thus, each iteration of peeling decoding produces a new \emph{residual} graph, indexed by the
iteration number~$\ell$.
Decoding is successful if the sequence of residual graphs leads to the empty graph, i.e., if all VNs
have been recovered. This occurs if at every iteration there is at least one degree-one CN. In contrast, decoding fails if there are no degree-one CNs left before reaching the empty graph.

Errors in the waterfall region mostly occur due to large (linear sized with respect to the component code
length) stopping sets~\cite{Amra09}.
The goal of the scaling law is, therefore, to estimate the probability that a linear-sized number of
VNs remains in the residual graph when peeling decoding stops.

For later use, we denote the frame, bit, and block error probability of a terminated ensemble
with coupling length $L$ decoded using full BP as
$P_{\mathsf{f,t}}^{(L)},$ $P_{\mathsf{b,t}}^{(L)}$, and $P_{\mathsf{bl,t}}^{(L)}$, respectively.
Likewise, we denote the frame, bit, and block error probability of an unterminated
ensemble (with error probability evaluated over the first $L'$ positions) as $P_{\mathsf{f,u}}^{(L')}$,
$P_{\mathsf{b,u}}^{(L')}$, and $P_{\mathsf{bl,u}}^{(L')}$.
Finally, the frame, bit, and block error probability of a terminated ensemble decoded using sliding
window decoding with window size $W$ is denoted as $P_{\mathsf{f,t,sw}}^{(L,W)}$, $P_{\mathsf{b,t,sw}}^{(L,W)}$,
and $P_{\mathsf{bl,t,sw}}^{(L,W)}$, respectively.

\subsection{Finite-Length Scaling of SC-LDPC Ensembles in~\cite{Olmo15}}
\label{sse:old}

\begin{figure}[!t]
    \centering
    \includegraphics[width=\columnwidth]{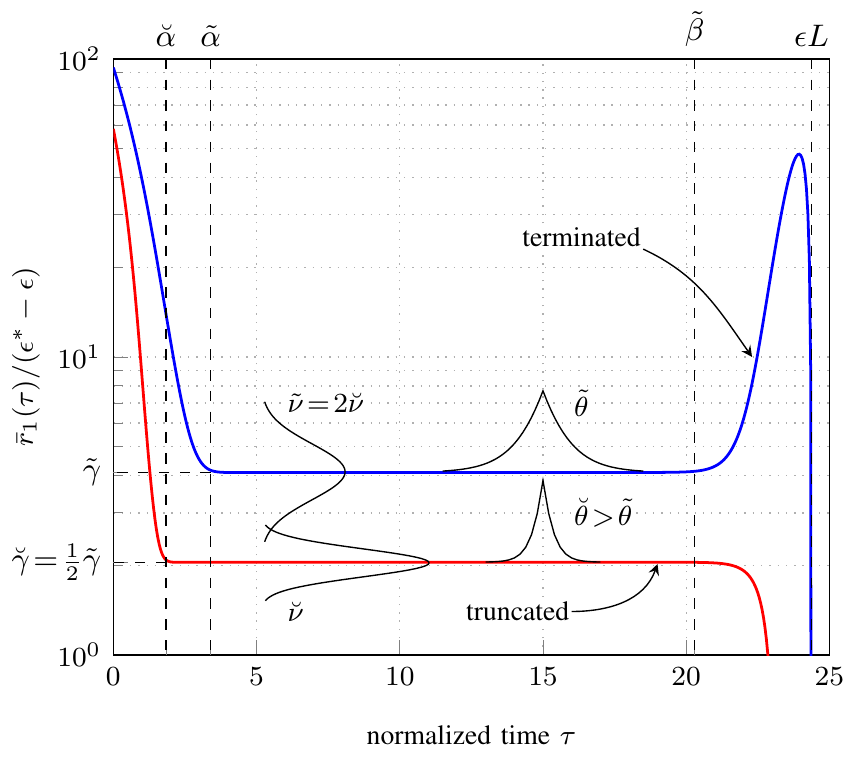}
    \vspace{-18pt}
    \caption{The evolution of the expected fraction of degree-one CNs $\rmean$ during peeling decoding,
    normalized by the distance to the BP threshold, for the
    $(5,10,L\!=\!50)$ ensemble with $\estar\!=\!0.4994$ at $\e\!=\!0.4875$.}
    \label{fig:parameters}
    \vspace{-4pt}
\end{figure}

Peeling decoding is successful if there is at least one degree-one CN at every iteration.
The number of degree-one CNs available for the peeling decoder throughout the iterations is thus a
crucial metric for estimating its performance. The scaling law in~\cite{Olmo15} is based on the stochastic process associated with the
fraction of degree-one CNs in the residual graphs~\cite{Luby01},
\begin{equation}
\rprocess \defn \frac{1}{N}\sum_u R_{1,u}(\tau),
\label{eq:rprocess}
\end{equation}
a quantity directly related to the number of degree-one CNs. In~\eqref{eq:rprocess}, $\tau \defn \ell / N$ can be viewed as the normalized time of the peeling decoding process, and
$R_{1,u}(\tau)$ is the number of degree-one CNs at position $u$ of the residual graph at
iteration~$\ell$.
The normalization of the decoding iteration and of the number of degree-one CNs by $N$
in~\eqref{eq:rprocess} allows to approximate $\rprocess$ by a
continuous-time real-valued stochastic process in the limit $N\grows$.
In~\cite{Amra09,Olmo15} it was shown that the distribution of $\rprocess$ for a fixed $\tau$
converges to a Gaussian distribution as $N\grows$.
Moreover, in the same limit $N\grows$ the realizations of $\rprocess$ concentrate around the
mean ${\rmean \defn \E{\rprocess}}$, with the expectation taken over the ensemble, channel, and
peeling decoding realizations.
Furthermore, it was noted in~\cite{Olmo15} that $\rmean$ exhibits a steady-state phase where it remains essentially constant.
We denote the range of $\tau$ corresponding to the steady state as~$\ssrange$.
As an illustration, $\rmean$ for the terminated $(5,10,L\!=\!50)$ SC-LDPC code ensemble at
\mbox{$\e\!=\!0.4875$} is shown in Fig.~\ref{fig:parameters} (blue curve).

As a decoding failure occurs if no degree-one CNs are available before all erased VNs have been recovered, to estimate the probability of
a decoding failure one needs to consider the normalized time of peeling decoding at which the number of degree-one CNs and hence
the value of $\rprocess$ drops to zero, referred to as the first hit time $\tau_0$,
\begin{equation}
    \tau_0 \defn \min \{\tau : \rprocess = 0 \}.
\end{equation}

Disregarding the normalized time $\e L$, at which
$\rmean$ drops to zero because the decoding process has performed as many iterations as the
average number of erased VNs to recover, the value of $\rmean$ is at its lowest during the steady state.
Decoding failures are therefore most likely to occur during the steady state. Hence, the scaling law in~\cite{Olmo15} assumes that $\tau_0 \in \ssrange$.
Consequently, the FER can be approximated as
\begin{equation}
    \FER \approx \int_{\sstart}^{\ssend} f_{\tau_0}(x) \der x,
\label{eq:fer}
\end{equation}
where $f_{\tau_0}$ denotes the PDF of $\tau_0$.

\tikzsetnextfilename{mean_evol_illustration}

Thus, estimating the FER requires the PDF $f_{\tau_0}$ for \mbox{$\tau_0 \in \ssrange$}.
This, in turn, requires studying the statistical properties of the decoding process $\rprocess$ for
$\tau \in \ssrange$, which are characterized by the following parameters.
\begin{enumerate}
    \item Expectation constant $\gamma$.
        The value of $\rmean$ in the steady state scales approximately linearly with the
        distance to the BP decoding threshold of the given $(\dv,\dc)$ SC-LDPC ensemble,
        $\estar$~\cite{Olmo15},
        \[
        \rmean \approx \gamma (\estar - \e),
        \]
        where $\epsilon$ is the channel erasure probability.
        Overall, the estimation of the expectation constant $\gamma$ requires running two relatively
        fast numerical procedures: 
        First, $\estar$ must be computed via density evolution~\cite{Lent10}.
        Second, the value of $\rmean$ in the steady state must be obtained by numerically
        solving a system of partial differential equations dubbed \textit{mean
        evolution}~\cite{Olmo15} for a channel parameter $\e$ sufficiently close to $\estar$
        and $L$ sufficiently large for the two decoding waves to form.
    \item Variance constant $\nu$.
        The variance of $\rprocess$ is shown to be inversely proportional to the component code
        length~$N$.
        A numerical solution to an augmented system of partial differential equations, called
        \textit{covariance evolution}, is required to estimate the variance of $\rprocess$
        in the steady state, which is modeled as independent of $\e$,
        \[
        \Var{\rprocess} \approx \frac{\nu}{N}.
        \]
    \item Correlation decay constant $\theta$.
        Finally, it is necessary to take into account the temporal correlation between different
        decoding iterations.
        For two time instants $\tau,\zeta \in \ssrange$, the covariance of the decoding process
        along iterations of peeling decoding is shown to be
        \[
        \E{\rprocess r_1(\zeta)} - \rmean \bar{r}_1(\zeta) \approx \frac{\nu}{N}\mathrm{e}^{-\theta
        \left| \zeta - \tau \right|}.
        \]
        The decay parameter $\theta$ is estimated in~\cite{Olmo15} from the solution to the
        covariance evolution using a semi-analytical technique that involves sampling from a
        multivariate Gaussian distribution and running mean evolution using these samples as
        initial conditions.
\end{enumerate}
In~\cite{Olmo15}, the equations for the mean and covariance evolution are
derived adapting the approach proposed in~\cite{Amra09} for uncoupled ensembles.

Overall, apart from the BP threshold $\estar$, the scaling law requires the five parameters
$(\sstart,\ssend,\gamma,\nu,\theta)$.
The meaning of these parameters is illustrated in Fig.~\ref{fig:parameters}.
The displayed relations between the parameters for the truncated and terminated ensembles have been confirmed by numerical solutions to mean evolution for $\gamma$ and by Monte-Carlo simulations for $\nu$ and $\theta$.
In the following, we denote the variables associated with the terminated ensembles with a tilde,
e.g., $\gammadouble$, and those associated with the truncated ensembles with a breve, e.g.,
$\gammasingle$.

The parameters $(\gammadouble,\nudouble,\thetadouble)$ are modeled in~\cite{Olmo15} as being dependent
only on $(\dv,\dc)$, i.e., as being independent of $\e,N,$ and $L$.
They were estimated for a channel parameter $\e = \estar - 0.04$.
The relatively significant margin is due to numerical stability issues that arise when solving
numerically the covariance evolution equations.
To avoid the dependency on $\e$, the range of the steady state $[\sstartdouble,\ssenddouble]$
is bounded as follows:
The fraction $\sstartlb$ of decoded bits of the uncoupled $(\dv,\dc)$-regular LDPC code ensemble
at the BP threshold~$\estar$ serves as a lower bound on $\sstartdouble$.
Likewise, the end of the steady state is upper bounded by $\ssenddouble = \e L$.

Assuming that the aforementioned approximations hold, the decoding process $\rprocess$ in the steady
state converges in the limit $N\grows$ to a stationary Gaussian Markov process with exponentially
decaying covariance.
The only stochastic process compatible with this description is an Ornstein-Uhlenbeck
process with appropriately chosen parameters, so $\rprocess$ is modeled in~\cite{Olmo15} by an
Ornstein-Uhlenbeck process.
Consequently, the distribution of $\tau_0$ in the steady state is approximated by the distribution
of the first hit time of the corresponding Ornstein-Uhlenbeck process.
It is known that the latter converges to an exponential distribution with mean $\m$
as $N \grows$,
\begin{equation}
    \m(\gamma,\nu,\theta) = \frac{\sqrt{2\pi}}{\theta}\int_0^{\gamma\sqrt{N/\nu}\left( \estar - \e \right)}
    \Phi(z)\mathrm{e}^{\frac{1}{2}z^2}\der z,
    \label{eq:olmos_mean}
\end{equation}
where $\Phi(z)$ is the cumulative distribution function (CDF) of the Gaussian distribution.
Thus, the PDF of $\tau_0$ in the steady state is approximated by an exponential PDF with scale
parameter $\m$, shifted by $\sstart$ to account for the initial transient period,
\begin{equation}
    f_{\tau_0}(x)\approx\fsingle(x) \defn \m^{-1} \exp\left(-\frac{x - \sstart}{\m}\right) H(x-\sstart),
    \label{eq:olmos_pdf_approx}
\end{equation}
where $H(x)$ is the Heaviside step function.

Finally, using the approximation~\eqref{eq:olmos_pdf_approx} in~\eqref{eq:fer} and the scaling
parameters $(\sstartlb,{\ssenddouble=\e L},\gammadouble,\nudouble,\thetadouble)$,
the FER of a terminated $(\dv,\dc,L,N)$ SC-LDPC code ensemble is estimated as~\cite{Olmo15}
\begin{equation}
P^{(L)}_{\mathsf{f}, \mathsf{t}, \text{\cite{Olmo15}}} 
\approx 1 - \exp\left(-\frac{\e L - \sstartlb}{\m(\gammadouble,\nudouble,\thetadouble)}\right).
\label{eq:olmos_fer}
\end{equation}

\section{Refined Scaling Law}
\label{sec:refinements}

The model in~\cite{Olmo15} assumes a single Ornstein-Uhlenbeck process.
However, the actual decoding process is characterized by two decoding waves.
Hence, in contrast to~\cite{Olmo15}, we propose modeling the decoding process $\rprocess$ in the
steady state of terminated SC-LDPC code ensembles as a combination of two identical and
independent Ornstein-Uhlenbeck processes, to better mimic this two-wave decoding.
Each Ornstein-Uhlenbeck process is the same as the equivalent process for the truncated ensemble,
where only one decoding wave is present.
The decoding is successful if the two decoding waves meet, otherwise decoding failure occurs.
The total number of iterations of peeling decoding is modeled as the sum of the first
hit times of the two component Ornstein-Uhlenbeck processes.
The crucial property of the proposed model is that these processes are allowed to fail (i.e.,
hit zero) independently.

\begin{figure}[!t]
    \centering
    \includegraphics[width=\columnwidth]{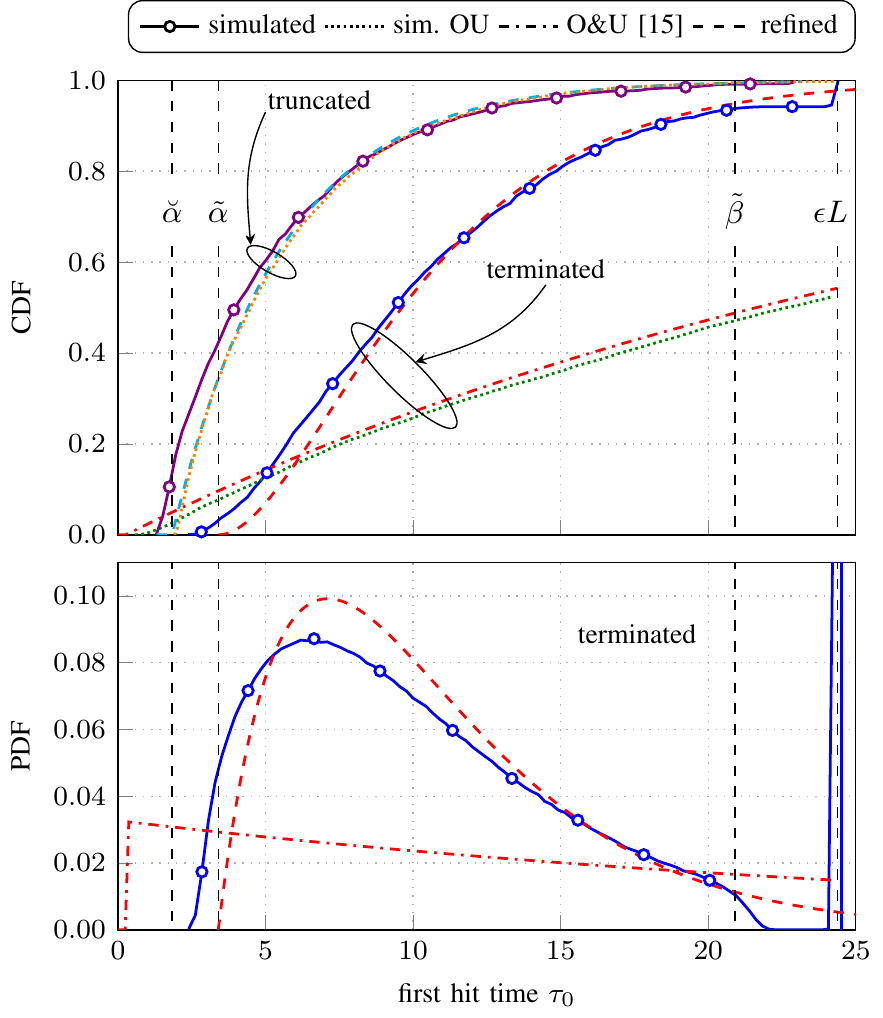}
    \vspace{-15pt}
    \caption{Comparison of the simulated and approximated CDFs of the first hit time for the terminated and
    truncated $(5,10,L\!=\!50,N\!=\!2000)$ ensembles at $\e = 0.4875$ (top). At the bottom plot,
    the corresponding PDFs for the terminated ensemble are shown.}
    \label{fig:cdf_5_10_04875}
    \vspace{-2ex}
\end{figure}

In the following, we provide a brief motivation for introducing this model.
In Fig.~\ref{fig:cdf_5_10_04875} (top) we compare the simulated CDF of $f_{\tau_0}$ (blue curve) with
the simulated CDF of the appropriately chosen Ornstein-Uhlenbeck process (green dotted curve) for
the terminated $(5,10,L\!=\!50,N\!=\!2000)$ \mbox{SC-LDPC} ensemble and \mbox{$\e=0.4875$}.
The corresponding analytical approximation $\fsingle(x)$ employed in~\cite{Olmo15} and given
in~\eqref{eq:olmos_pdf_approx} is plotted as the red dash-dotted curve.
The corresponding PDFs are shown in Fig.~\ref{fig:cdf_5_10_04875} (bottom).
A significant disagreement between the distributions of the first hit time of the Ornstein-Uhlenbeck
process and that of the first hit time of peeling decoding indicates that a single
Ornstein-Uhlenbeck process is inadequate as a model for the peeling decoding process $\rprocess$ in
the steady state.
In particular, it is clear from Fig.~\ref{fig:cdf_5_10_04875} (bottom) that the simulated
$f_{\tau_0}$ does not follow the exponential distribution.
On the other hand, the first hit time of the Ornstein-Uhlenbeck process is well approximated by
the exponential distribution, as indicated by the agreement between the red dash-dotted and green
dotted curves.

In the figure, we also plot the simulated CDF of the first hit time of peeling decoding (purple
curve), the simulated CDF of the first hit time of the Ornstein-Uhlenbeck process (orange dotted
curve) and its exponential approximation (cyan dashed curve) for the corresponding truncated
ensemble.
The match between these three curves is much closer than for the case of the terminated ensemble.
Hence, we conclude that in the case where a single decoding wave is present, a single
Ornstein-Uhlenbeck process is an adequate model for the decoding process, whereas two-wave
decoding requires a refined model, motivating the proposed model based on two Ornstein-Uhlenbeck
processes for the terminated case.

\tikzsetnextfilename{cdf_term_5_10_04875}

\subsection{Decoding Process as Two Independent Ornstein-Uhlenbeck Processes}
\label{sse:two_waves_approx}
Assuming that the individual Ornstein-Uhlenbeck processes may fail independently, we model the first
hit time of the combined decoding process as the sum of the two first hit times of the component
processes.
As the exponential distribution yields a good approximation of the first hit time for the
single-wave decoding, we approximate the PDF of the first hit time of the decoding process for the
terminated ensemble as the convolution of two exponential PDFs, or, equivalently, as the PDF of the
Erlang distribution with shape parameter $2$ and scale parameter $\m$,
\begin{equation}
    \fdouble(x) \defn \m^{-2}\left(x-\sstart\right)\exp\left(-\frac{x-\sstart}{\m}\right)H(x-\sstart),
    \label{eq:terminated_pdf_approx}
\end{equation}
where $\m$ is given in~\eqref{eq:olmos_mean}.
As in~\eqref{eq:olmos_pdf_approx}, we shift the PDF of the Erlang distribution by
$\sstart$ to the beginning of the steady-state regime.

Thus, for the terminated ensemble, we approximate the PDF of $\tau_0$ in the steady state as
\begin{equation}
\label{eq:approx2}
f_{\tau_0}(x)\approx\fdouble(x).
\end{equation}
Using \eqref{eq:terminated_pdf_approx}--\eqref{eq:approx2} in~\eqref{eq:fer}, the FER of the
terminated SC-LDPC code ensemble can then be approximated as
\begin{equation}
    \FERt^{(L)} \approx 1 - \left( 1 + \frac{\ssenddouble - \sstartdouble}{\m(\gammasingle,\nusingle,\thetasingle)} \right)
    \exp \left( - \frac{\ssenddouble - \sstartdouble}{\m(\gammasingle,\nusingle,\thetasingle)} \right).
    \label{eq:our_fer}
\end{equation}
We emphasize that the triple $(\gamma,\nu,\theta)$ in the refined scaling law~\eqref{eq:our_fer}
corresponds to the propagation of a \textit{single} decoding wave and should therefore be estimated
from the truncated ensemble.
The pair $(\sstart,\ssend)$, on the other hand, must be estimated from the terminated ensemble,
since these parameters determine the boundaries of the two-wave regime that we are ultimately
interested in.
To summarize, the FER in~\eqref{eq:our_fer} should be evaluated using the scaling parameters
$(\sstartdouble,\ssenddouble,\gammasingle,\nusingle,\thetasingle)$.

It is important to remark that the authors of~\cite{Olmo15} mentioned that considering the
decoding process as two processes corresponding to the two decoding waves would affect the scaling
constants $\nu,\gamma$, and $\theta$, from which the constants for the combined process (still with
the first hit time modeled as an exponential distribution) could be obtained.
Here, however, we argue that the separate treatment of the two decoding waves leads not only to a
change in the scaling constants, but also to the change of the distribution of the first hit time of
the combined process from an exponential to an Erlang distribution.

\subsection{Dependence of the Scaling Parameters on the Channel Parameter}
\label{sse:additional_refinements}

The scaling parameters $\sstart,\gamma,\nu,$ and $\theta$ are modeled in~\cite{Olmo15}
as constants independent of the channel parameter~$\e$.
Their estimation for different values of $\e$, however, yields different
numerical values, which indicates that they are, in fact, dependent on $\e$.
Therefore, it is preferable to treat these parameters as functions of $\e$.
For some of the scaling parameters this approach is feasible.
In particular, the triple $(\sstart,\ssend,\gamma)$ can be obtained in a reasonable time from the mean
evolution $\rmean$~\cite{Olmo15}.
We note that the boundaries of the steady state $\ssrange$ depend also on the length of the coupled
chain $L$.
For each $(\dv,\dc,L)$ SC-LDPC ensemble, we estimate $\sstart,\ssend,$ and $\gamma$ from the
evolution of $\rmean$ for a number of channel parameters $\e$ and obtain the intermediate values by
linear interpolation.

Treating $\nu$ and $\theta$ as functions of $\e$, on the other hand, is impractical, since it
requires numerically solving the significantly more complex covariance evolution for each value of
the channel parameter $\e$, which renders the approach infeasible.
We thus follow~\cite{Olmo15} and model~$\nu$ and~$\theta$ as independent of~$\e$.
In this work, we estimate these two constants via Monte-Carlo simulations of peeling decoding
by setting $N=10^4$ and choosing the highest $\e$ for which the system operates in an
effectively error-free regime.
For our running example of the $(5,10,L)$ SC-LDPC code ensemble, these parameters are estimated at $\e = 0.485$ as $\nusingle \approx 0.424$ and~$\thetasingle \approx 1.64$.

The red dashed lines in Fig.~\ref{fig:cdf_5_10_04875} correspond to the CDF and the PDF of the
employed Erlang approximation~\eqref{eq:approx2} with $\sstartdouble$, $\ssenddouble$, and
$\gammasingle$ computed for $\epsilon= 0.4875$.
The good agreement of the approximation with the simulated distribution of the first hit time of
peeling decoding supports the proposed model.

\subsection{Scaling Law to Predict the Bit Error Rate}

The above-described scaling law for the FER can be easily extended to predict the BER of a
terminated SC-LDPC code ensemble, i.e., the fraction of bits that remain erased when decoding terminates.
Suppose that the peeling decoder halted at normalized time $\tau_0\!=\!x$.
In that case it would have performed $xN$ decoding iterations before the failure and hence
approximately \mbox{$\e LN-xN$} out of $LN$ bits would remain unrecovered.
Accordingly, the BER can be approximated by averaging the fraction of undecoded bits over the
distribution of the first hit time,
\begin{equation}
\label{eq:ber}
    \BER \approx \int_{\sstart}^{\ssend}  \left(\e - \frac{x}{L}\right) f_{\tau_0}(x) \der x.
\end{equation}
The BER performance of the terminated
ensemble can then be predicted by using the approximation~\eqref{eq:terminated_pdf_approx}--\eqref{eq:approx2}
in~\eqref{eq:ber} with parameters $(\sstartdouble,\ssenddouble,\gammasingle,\nusingle,\thetasingle)$
as
\begin{align}
    \BERt^{(L)} &\approx \frac{\e L - \sstartdouble - 2\mb}{L} \label{eq:our_ber}\\
    & +\exp \left(\frac{\sstartdouble-\ssenddouble}{\mb}\right) \frac{\ssenddouble^2 + \sstartdouble\e L - \left( \e L +
    \sstartdouble - 2\mb \right) \left( \ssenddouble + \mb \right)}{\mb L}, \nonumber
\end{align}
where $\mb=\m(\gammasingle,\nusingle,\thetasingle)$.

\subsection{Scaling Law to Predict the Block Error Rate}
\label{sse:bler}
A similar extension of the scaling law allows us to predict the BLER of a terminated SC-LDPC code ensemble.
Again, suppose the decoder performed $xN$ iterations before halting.
The number of spatial positions (or blocks) containing erased VNs depends on the speed with which the decoding waves
propagate through the coupled chain.
Let us assume the waves traverse $\speed$ positions in $N$ iterations.
Then approximately $(x - \sstart) \speed$ out of $L$ blocks would be free of erased VNs.
Additionally, successful decoding implies that all $L$ blocks are decoded correctly.
Putting it all together and averaging over the distribution of the first hit time, we obtain
\begin{align}
    \BLER &\approx 1 - \frac{1}{L} \left(\int_{\sstart}^{\ssend} (x-\sstart)\speed f_{\tau_0}(x) \der x + (1 - \FER) L\right)  \nonumber\\
    &= \FER - \frac{\speed}{L} \int_{\sstart}^{\ssend} (x-\sstart) f_{\tau_0}(x) \der x. \label{eq:bler}
\end{align}

For the terminated ensemble, the approximation~\eqref{eq:terminated_pdf_approx}--\eqref{eq:approx2} should be used
in~\eqref{eq:bler} with parameters $(\sstartdouble,\ssenddouble,\gammasingle,\nusingle,\thetasingle)$,
resulting in
\begin{equation}
    \BLERt^{(L)} \approx \FERt^{(L)} - \frac{\speed\mb}{L} \Bigg(\exp \left( -\xi \right)
    (\xi^2 + 2\xi + 2) + 2 \Bigg),
    \label{eq:our_bler}
\end{equation}
where $\xi = (\ssenddouble - \sstartdouble)/\mb$.

It remains to show how to estimate $\speed$, the speed of the decoding waves.
Since every iteration of the peeling decoder recovers exactly one VN, it will take the waves as many iterations to
propagate by one position as there are erased VNs in one position to decode.
Accordingly, we estimate $\speed$ from the average number of VNs in the middle of the coupled chain during the steady
state as
\begin{equation}
    \label{eq:calc_speed}
    \speed \approx N \E{V_{\lfloor L/2 \rfloor} \left(\frac{\ssend - \sstart}{2}\right)}^{-1},
\end{equation}
where $\E{V_u(\tau)}$, the average number of VNs at position $u$ at normalized iteration $\tau$, is produced alongside
$\rmean$ by numerically solving mean evolution.
As with the other parameters that we estimate from mean evolution, namely, $\sstart, \ssend,$ and $\gamma$, we treat $\speed$
as a function of $\e$ by evaluating~\eqref{eq:calc_speed} for a number of channel parameters and linearly interpolating
values in between.

To summarize, the refined prediction of the FER, BER, and BLER performance of the terminated $(\dv,\dc,L,N)$
SC-LDPC code ensemble is given by~\eqref{eq:our_fer},~\eqref{eq:our_ber}, and~\eqref{eq:our_bler}, respectively, with
parameters
$(\sstartdouble_\e,\ssenddouble_\e,\gammasingle_\e,\speede,\nusingle,\thetasingle)$.
The dependence of the parameters on $\epsilon$ is highlighted with the subscript $\e$.

Note that the work \cite{Olmo15} did not consider a scaling law for the BER or BLER.
However, the FER scaling law in~\cite{Olmo15} can also be extended to the BER and BLER following the same
reasoning as outlined above.

\subsection{Scaling Law for the Unterminated SC-LDPC Code Ensemble}
\label{sec:ScalingUnterminated}

For the finite-length scaling analysis of SC-LDPC code ensembles under window decoding, addressed in the next section, we will require the scaling law for the unterminated ensemble, with error probability evaluated over $L'$ spatial positions. As the unterminated ensemble is characterized by a single decoding wave, the finite-length scaling proposed in \cite{Olmo15}, which considers a single process (see \eqref{eq:olmos_fer}), can be used with appropriate choice of the parameters $(\alpha,\beta,\gamma,\nu,\theta)$. In particular, the parameters $(\gamma,\nu,\theta)$ of the unterminated ensemble are identical to those of the corresponding truncated ensemble, i.e., $(\gammasingle,\nusingle,\thetasingle)$. Furthermore, we set $\beta=\epsilon L'$, which results from the fact that we need to contemplate the propagation of the decoding wave only until position $L'$ but no truncation occurs after the first $L'$ positions (the chain is semi-infinite).
Finally, the beginning of the steady state $\sstart$ for the unterminated ensemble depends on the
schedule employed; we assume that only degree-one CNs from the first $L'$ positions are removed
at the initial phase (i.e., before the wave is formed),
which is equivalent to considering the truncated ensemble.
Accordingly, we set $\sstart$ to that of an ensemble truncated after $L'$ positions.

Thus, the frame error rate of an unterminated SC-LDPC code ensemble evaluated over $L'$ spatial positions can be approximated as 
\begin{align}
\FERu^{(L')} \approx 1 - \exp\left(-\frac{\epsilon L' - \sstartsingle}{\mb}\right).
\label{eq:untF}
\end{align}
Similarly, using \eqref{eq:olmos_pdf_approx} in \eqref{eq:ber} with parameters $(\sstartsingle,\beta=\e L',\gammasingle,\nusingle,\thetasingle)$, the bit error rate can be written as
\begin{align}
\BERu^{(L')} \approx \frac{\mb}{L'}\cdot\exp\left(-\frac{\e
L'-\sstartsingle}{\mb}\right)+\frac{\e L'-\sstartsingle-\mb}{L'}.
\label{eq:untB}
\end{align}
Finally, we could use~\eqref{eq:olmos_pdf_approx} in~\eqref{eq:bler} to estimate the block error rate of an
unterminated ensemble.
Instead, we propose a slightly more accurate approximation.
Suppose the wave propagated through $2.5$ spatial positions before decoding failure.
Then just $2$ blocks (not $2.5$) would be decoded successfully.
The estimation~\eqref{eq:bler} disregards this and is therefore too optimistic.
Improving the estimation for the terminated ensemble requires considering each of the two waves separately, which complicates the derivations.
In the case of the unterminated ensemble, however, where only one decoding wave is present, this effect is easy to take into
account. We can estimate the block error rate as
\begin{align}
    \label{eq:bler_floor}
    \BLER &\approx \FER - \frac{1}{L'} \int_{\sstart}^{\ssend} \lfloor (x-\sstart)\speed \rfloor f_{\tau_0}(x) \der x \nonumber\\
    &= \FER - \frac{1}{L'} \sum_{i=0}^{\lfloor (\ssend - \sstart) \speed \rfloor - 1} i \int_{i/\speed}^{(i+1)/\speed}
    f_{\tau_0}(x+\sstart) \der x \nonumber\\
    &- \frac{\lfloor (\ssend - \sstart) \speed \rfloor}{L'}
    \int_{\lfloor (\ssend - \sstart) \speed \rfloor / \speed}^{\ssend - \sstart} f_{\tau_0} (x + \sstart) \der x.
\end{align}
Using~\eqref{eq:olmos_pdf_approx} in~\eqref{eq:bler_floor}
with parameters $(\sstartsingle,\beta=\e L',\speed,\gammasingle,\nusingle,\thetasingle)$,
we obtain the estimation of the block error rate as
\begin{align}
    \label{eq:bler_unterm}
    \BLERu^{(L')} &\approx \FERu^{(L')} \\
    &- \frac{1}{L'} \sum_{i=0}^{\lfloor \omega \speed \rfloor - 1}
    i \Bigg[ \exp\left( -\frac{i}{\speed\mb} \right) - \exp \left( -\frac{i + 1}{\speed\mb} \right) \Bigg] \nonumber\\
    &- \frac{\lfloor \omega \speed \rfloor}{L'}
    \Bigg[ \exp \left( -\frac{\lfloor \omega \speed \rfloor}{\speed\mb}\right)
    - \exp \left( -\frac{\omega}{\mb} \right) \Bigg],\nonumber
\end{align}
where $\omega = \e L' - \sstartsingle$.

The parameter $\speed$ should be estimated using~\eqref{eq:calc_speed}.
The considerations in Section~\ref{sse:bler} apply to both terminated and unterminated ensembles.

\section{Finite-Length Scaling of SC-LDPC Codes Under Window Decoding}

In this section, we extend the finite-length scaling derived in the previous section to predict the finite-length performance of SC-LDPC code ensembles under window decoding, i.e., we address the scaling for the decoding approach used in practice.
In particular, we consider a window size $W$ and, as before, a terminated coupled chain of $L$
positions.

\begin{figure}
    \centering
    \includegraphics[width=\columnwidth]{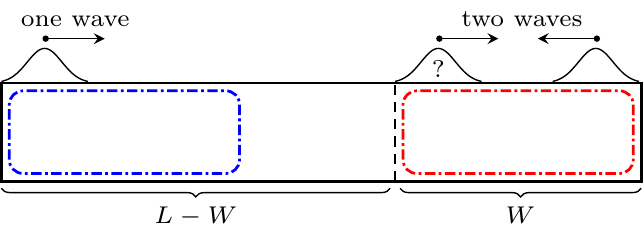}
    \vspace{-15pt}
    \caption{Schematic representation of the two-phase decoding process corresponding to the decoding of terminated SC-LDPC codes under window decoding with window size $W$.
    The initial and final positions of the sliding window are shown in blue and red, respectively.}
    \label{fig:protograph_window}
    \vspace{-2ex}
\end{figure}

The proposed finite-length scaling is based on the observation that, as briefly discussed in Section~\ref{sec:preliminaries}, the decoding of a terminated
SC-LDPC code under window decoding is characterized by two different phases.
In the first phase, a single decoding wave propagates from the beginning of the chain inward along
$L-W$ coupled positions.
This corresponds to the sliding of the decoding window from the beginning of the chain until the
window comprises positions in the range $[L-2W+1,L-W]$.
If the decoding wave propagates until position $L-W$, the decoding of the last $W$ positions
(corresponding to the case when the window reaches the end of the coupled chain) is then
characterized by two decoding waves that propagate inward from the boundaries of the window. This two-phase process is schematized in Fig.~\ref{fig:protograph_window}. Otherwise, if the decoding wave of the first phase does not propagate until the $L-W$ position,
i.e., it stops earlier, the decoding of the last $W$ positions is characterized by a single decoding
wave that propagates inward from the right-termination of the chain.

Following this observation, we model the decoding of SC-LDPC code ensembles under window decoding as
a two-phase decoding process, where the first phase corresponds to the first $L-W$ positions of the
coupled chain and the second phase to the last $W$ positions.

\subsection{Frame Error Probability}

Decoding is successful if, in the first phase, the (single) decoding wave reaches position $L-W$, and in the second phase the two decoding waves propagating along the last $W$ positions meet.
Denote by $\Pfone$ the frame error probability of the first phase and by $\Ptftwo$ the frame error probability of the second phase given that the decoding wave of the first phase has successfully propagated to position $L-W$,
\begin{equation}
\Ptftwo \defn \operatorname{Pr}\left\{ \text{error in 2nd phase\,} \big| \text{\,success in 1st phase} \right\}.
    \label{eq:pftwo_def}
\end{equation}
The frame error probability of a terminated SC-LDPC code ensemble under sliding window
decoding with window size $W$ can then be written as
\begin{align}
\label{eq:PfSW}
P_{\mathsf{f,t,sw}}^{(L,W)} = 1-\left(1-\Pfone\right)\left(1-\Ptftwo\right),
\end{align}
which follows from the law of total probability.

Note that the decoding of the first $L-W$ positions of the coupled chain (i.e., the first phase) corresponds to the decoding of an unterminated SC-LDPC code ensemble where the error probability must be evaluated over the $L'=L-W$ first positions. Thus,
\begin{align}
\label{eq:Pfone}
\Pfone = P_{\mathsf{f,u}}^{(L-W)}.
\end{align}

If the decoding
wave of the first phase propagates until position $L-W$, the decoding of the last $W$ positions of the coupled chain corresponds to the decoding of a terminated SC-LDPC code ensemble of chain length $W$. Hence,
\begin{align}
\label{eq:Ptftwo}
\Ptftwo = P_{\mathsf{f,t}}^{(W)}.
\end{align}

Finally, using \eqref{eq:Pfone} and \eqref{eq:Ptftwo} in \eqref{eq:PfSW}, the finite-length scaling becomes
\begin{align}
\label{eq:PfSWb}
P_{\mathsf{f,t,sw}}^{(L,W)} = 1-\left(1-P_{\mathsf{f,u}}^{(L-W)}\right)\left(1-P_{\mathsf{f,t}}^{(W)}\right),
\end{align}
where $P_{\mathsf{f,t}}^{(\cdot)}$ is given in \eqref{eq:our_fer}, and $P_{\mathsf{f,u}}^{(\cdot)}$ is given in \eqref{eq:untF}.
Let us emphasize that in the second term of the product in~\eqref{eq:PfSWb} we may assume the presence of two decoding waves (and hence the terminated ensemble) because it represents the conditional probability of successful decoding in the second phase given successful decoding in the first, which implies the presence of the wave from the left boundary of the chain.

\subsection{Bit Error Probability}
\label{sse:wd_ber}

Let $\Pbone$ and $\Pbtwo$ denote the bit error probability of the first $L-W$ positions and the last $W$ positions of the coupled chain, respectively.
The bit error probability of a terminated SC-LDPC code ensemble under sliding window
decoding can be obtained as a linear combination of $\Pbone$ and $\Pbtwo$. In particular, the
fraction of coded bits corresponding to the first $L-W$ coupled positions is
$\frac{(L-W)N}{LN}=1-\frac{W}{L}$ and the fraction of coded bits corresponding to the
last $W$ positions is \mbox{$\frac{WN}{LN}=\frac{W}{L}$}.
Thus, the bit error probability can be written as
\begin{align}
\label{eq:PbSW}
P_{\mathsf{b,t,sw}}^{(L,W)} = \Pbone\cdot\left(1-\frac{W}{L}\right)+\Pbtwo \cdot \frac{W}{L}.
\end{align}
Following the same reasoning as for the frame error rate, $\Pbone$ corresponds to the bit error probability of an unterminated SC-LDPC code ensemble with the error probability evaluated over the $L'=L-W$ first positions, i.e., 
\begin{align}
\label{eq:Pbone}
\Pbone = P_{\mathsf{b,u}}^{(L-W)}.
\end{align}

On the other hand, $\Pbtwo$ depends on whether the decoding wave of the first phase propagates until
position $L-W$ or not. In the first case, which occurs with probability
$1-\Pfone=1-P_{\mathsf{f,u}}^{(L-W)}$, the bit error probability of the last $W$ positions
corresponds to the bit error probability of a terminated SC-LDPC of length $W$, i.e.,
$P_{\mathsf{b,t}}^{(W)}$. In the second case, which occurs with probability
$\Pfone=P_{\mathsf{f,u}}^{(L-W)}$, the bit error probability of the last $W$ positions corresponds to the bit error probability of an unterminated SC-LDPC ensemble (terminated from the right but unterminated from the left) with the error probability evaluated over $L'=W$ positions. Thus,
\begin{align}
\label{eq:Pbtwo}
\Pbtwo = P_{\mathsf{b,t}}^{(W)}\cdot\left(1-P_{\mathsf{f,u}}^{(L-W)}\right) + P_{\mathsf{b,u}}^{(W)}\cdot P_{\mathsf{f,u}}^{(L-W)}.
\end{align}

Using \eqref{eq:Pbone} and \eqref{eq:Pbtwo} in \eqref{eq:PbSW}, we obtain
\begin{align}
\label{eq:PbSWb}
P_{\mathsf{b,t,sw}}^{(L,W)} &= P_{\mathsf{b,u}}^{(L-W)}\cdot\left(1-\frac{W}{L}\right)\\
&+\left(P_{\mathsf{b,t}}^{(W)}\cdot\left(1-P_{\mathsf{f,u}}^{(L-W)}\right) + P_{\mathsf{b,u}}^{(W)}\cdot
P_{\mathsf{f,u}}^{(L-W)}\right) \cdot\frac{W}{L},\nonumber
\end{align}
where $P_{\mathsf{b,t}}^{(\cdot)}$ is given in \eqref{eq:our_ber} and $P_{\mathsf{b,u}}^{(\cdot)}$ is given in \eqref{eq:untB}.

\subsection{Block Error Probability}
Let us denote the block error probabilities of the first and second phase by $\Pblone$ and $\Pbltwo,$ respectively.
Similarly to the calculation of the BER in Section~\ref{sse:wd_ber}, we obtain the estimation of the BLER as
\begin{align}
\label{eq:PbSWbl}
P_{\mathsf{bl,t,sw}}^{(L,W)} &= \Pblone\cdot\left(1-\frac{W}{L}\right)+\Pbltwo \cdot \frac{W}{L}\\
&=P_{\mathsf{bl,u}}^{(L-W)}\cdot\left(1-\frac{W}{L}\right) \nonumber\\
&+\left(P_{\mathsf{bl,t}}^{(W)}\cdot\left(1-P_{\mathsf{f,u}}^{(L-W)}\right) + P_{\mathsf{bl,u}}^{(W)}\cdot
P_{\mathsf{f,u}}^{(L-W)}\right) \cdot\frac{W}{L}, \nonumber
\end{align}
where $\BLERt^{(\cdot)}$ and $\BLERu^{(\cdot)}$ are given in \eqref{eq:our_bler} and \eqref{eq:bler_unterm}, respectively.

We remark that the finite-length scaling in \eqref{eq:PfSWb}, \eqref{eq:PbSWb}, and \eqref{eq:PbSWbl} assumes that the
window size $W$ is big enough for the decoding process to reach the steady state, i.e., for a decoding wave to be formed.
As shown in the numerical results section, the scaling law is very
accurate for window sizes $W\ge 10$.
It is also worth mentioning that the window decoding algorithm proposed in~\cite{Iye12} and used in our numerical
simulations delays the decision on a bit by $\dv - 1$ additional VN positions.
The simplified two-phase model introduced in this section does not take that into account.
Finally, note that for large coupled chains the error probability
will be dominated by the error probability of the first phase, i.e.,
$\Pfone\gg \Ptftwo,$ $\Pbone\gg \Pbtwo,$ and $\Pblone \gg \Pbltwo$.

\section{Numerical Results}
\label{sec:numerical-results}

\tikzsetnextfilename{our_approx_5_10_term}
\begin{figure}[!t]
    \centering
    \includegraphics[width=\columnwidth]{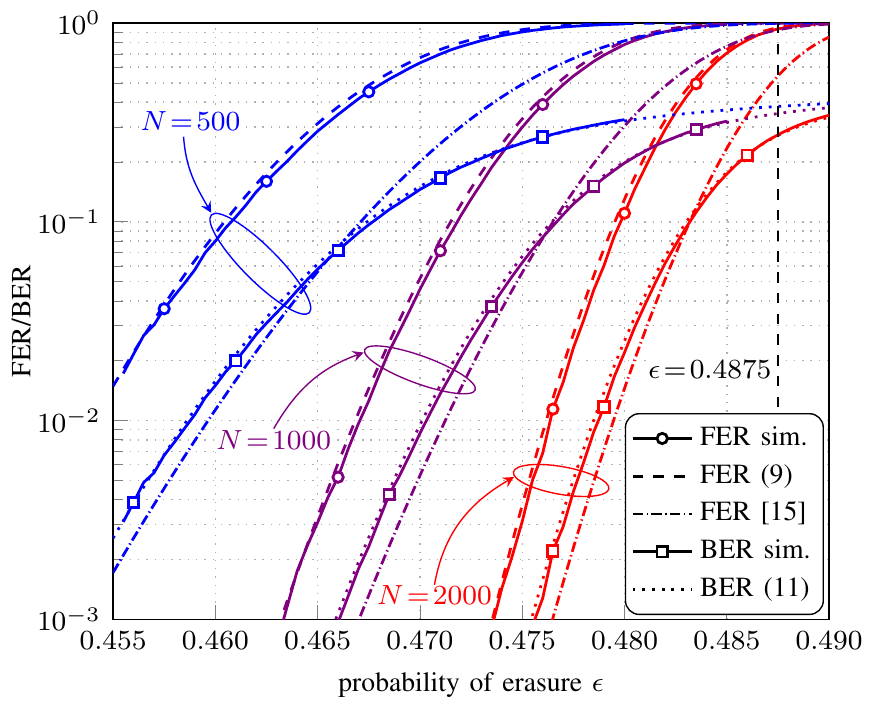}
    \vspace{-15pt}
    \caption{Simulated FER and BER curves and the corresponding analytical approximations
    for the terminated $(5,10,L\!=\!50,N)$ ensemble with \mbox{$\estar\!=\!0.4994$} for different $N$.
    The parameters $\nusingle$ and $\thetasingle$ are estimated at $\e = 0.485$ to be
    $\nusingle \approx 0.424$ and $\thetasingle \approx 1.64$.}
    \label{fig:our_approx_5_10}
    \vspace{-2ex}
\end{figure}

\tikzsetnextfilename{refinements}
\begin{figure}[!t]
    \centering
    \includegraphics[width=\columnwidth]{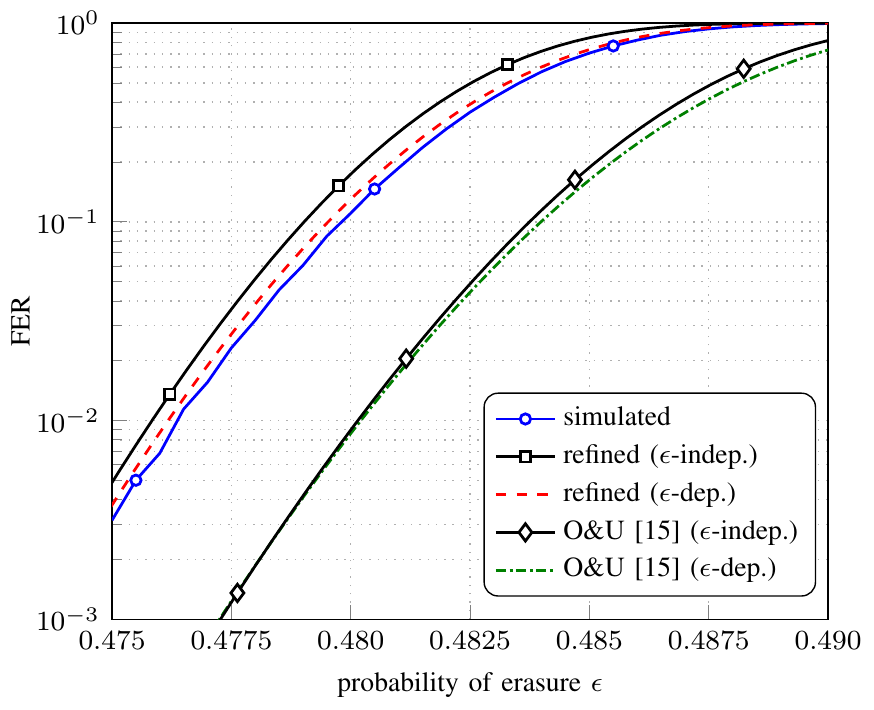}
    \vspace{-15pt}
    \caption{The effect of introducing the dependency of $(\sstart,\ssend,\gamma)$ on the channel parameter $\e$ to predict the FER of the terminated $(5,10,L\!=\!50,N\!=\!2000)$ ensemble under full BP decoding (blue line with circles).}
    \label{fig:refinements}
    \vspace{-2ex}
\end{figure}

\tikzsetnextfilename{our_approx_4_8_term}
\begin{figure}[!t]
    \centering
    \includegraphics[width=\columnwidth]{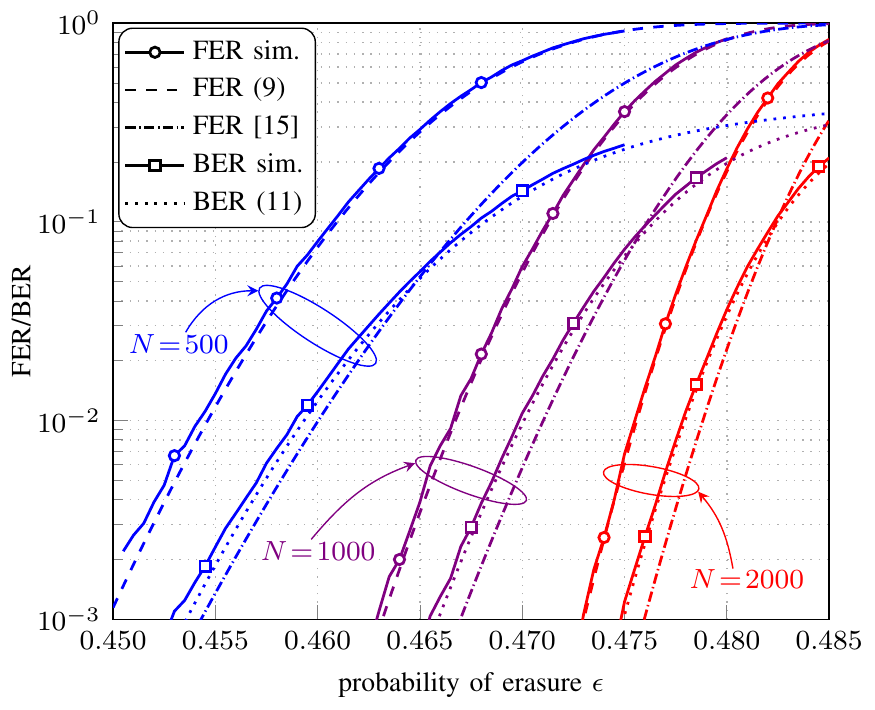}
    \vspace{-15pt}
    \caption{Simulated FER and BER curves and the corresponding analytical approximations
    for the terminated $(4,8,L\!=\!50,N)$ ensemble with \mbox{$\estar\!=\!0.4977$} for different $N$.
    The parameters $\nusingle$ and $\thetasingle$ are estimated at $\e = 0.48$ to be
    $\nusingle \approx 0.406$ and $\thetasingle \approx 1.47$.}
    \label{fig:our_approx_4_8_term}
    \vspace{-2ex}
\end{figure}

\tikzsetnextfilename{our_approx_3_6_term}
\begin{figure}[!t]
    \centering
    \includegraphics[width=\columnwidth]{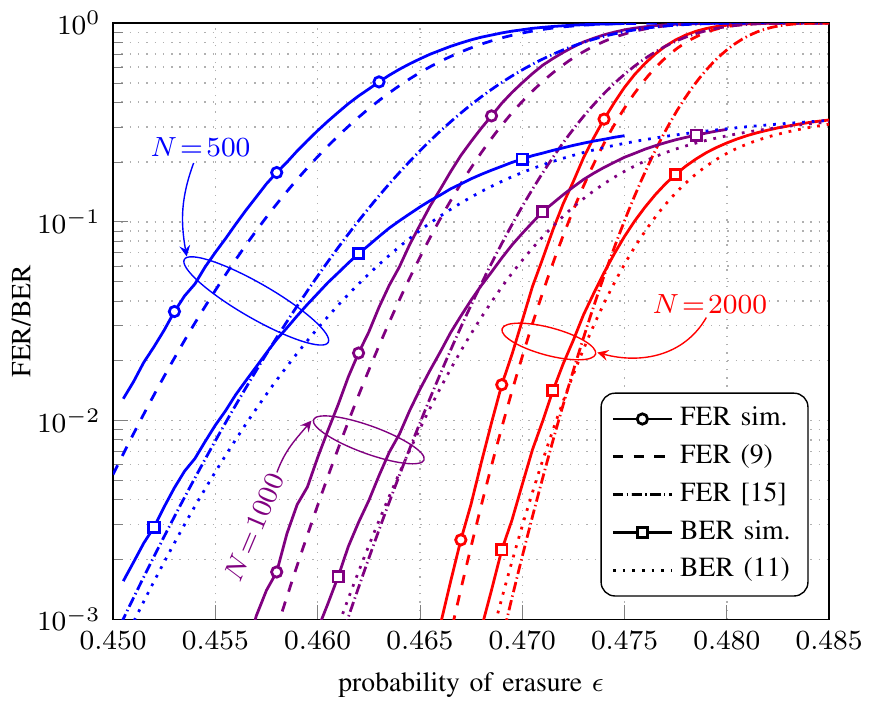}
    \vspace{-15pt}
    \caption{Simulated FER and BER curves and the corresponding analytical approximations
    for the terminated $(3,6,L\!=\!50,N)$ ensemble with \mbox{$\estar\!=\!0.4881$} for different $N$.
    The parameters $\nusingle$ and $\thetasingle$ are estimated at $\e = 0.475$ to be
    $\nusingle \approx 0.338$ and $\thetasingle \approx 1.28$.}
    \label{fig:our_approx_3_6_term}
    \vspace{-2ex}
\end{figure}

\tikzsetnextfilename{bec_5_10_sw_fer}
\begin{figure}[!t]
    \centering
    \includegraphics[width=\columnwidth]{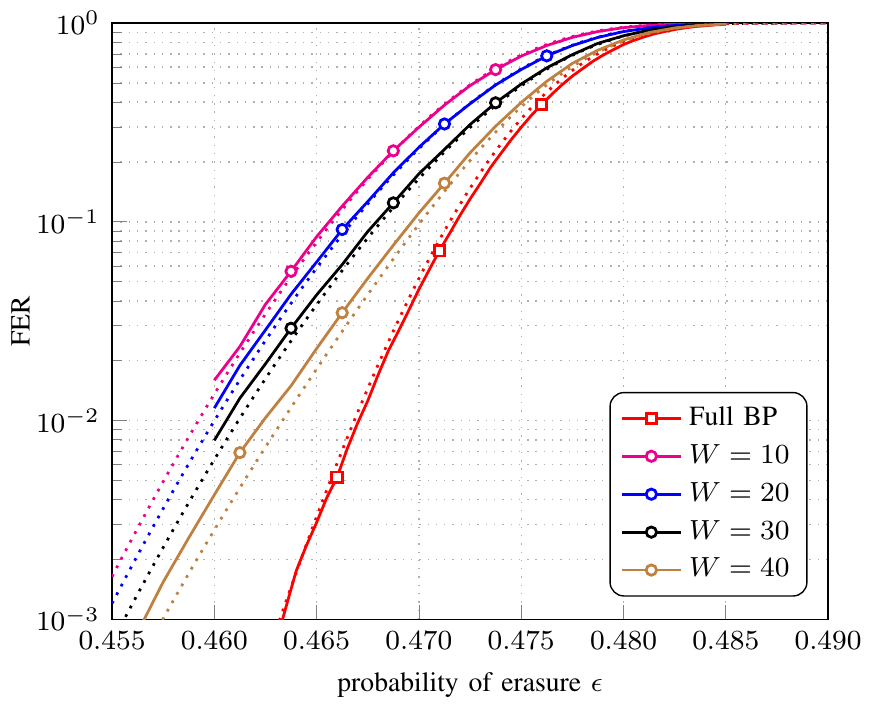}
    \vspace{-15pt}
    \caption{Simulated FER curves (solid lines) and corresponding analytical approximations (dotted
    lines) for the terminated $(5,10,L\!=\!50,N\!=\!1000)$ ensemble for different values of the window size $W$.}
    \label{fig:FER_5_10}
    \vspace{-2ex}
\end{figure}

\tikzsetnextfilename{bec_5_10_sw_ber}
\begin{figure}[!t]
    \centering
    \includegraphics[width=\columnwidth]{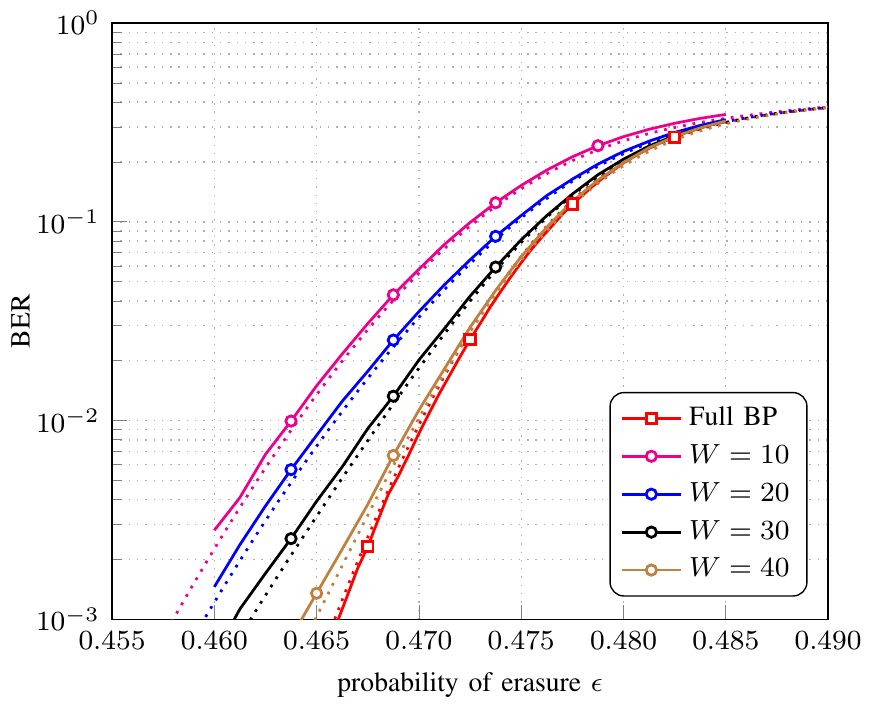}
    \vspace{-15pt}
    \caption{Simulated BER curves (solid lines) and corresponding analytical approximations (dotted
    lines) for the terminated $(5,10,L\!=\!50,N\!=\!1000)$ ensemble for different values of the window size $W$.}
    \label{fig:BER_5_10}
    \vspace{-2ex}
\end{figure}

\tikzsetnextfilename{bec_5_10_sw_bler}
\begin{figure}[!t]
    \centering
    \includegraphics[width=\columnwidth]{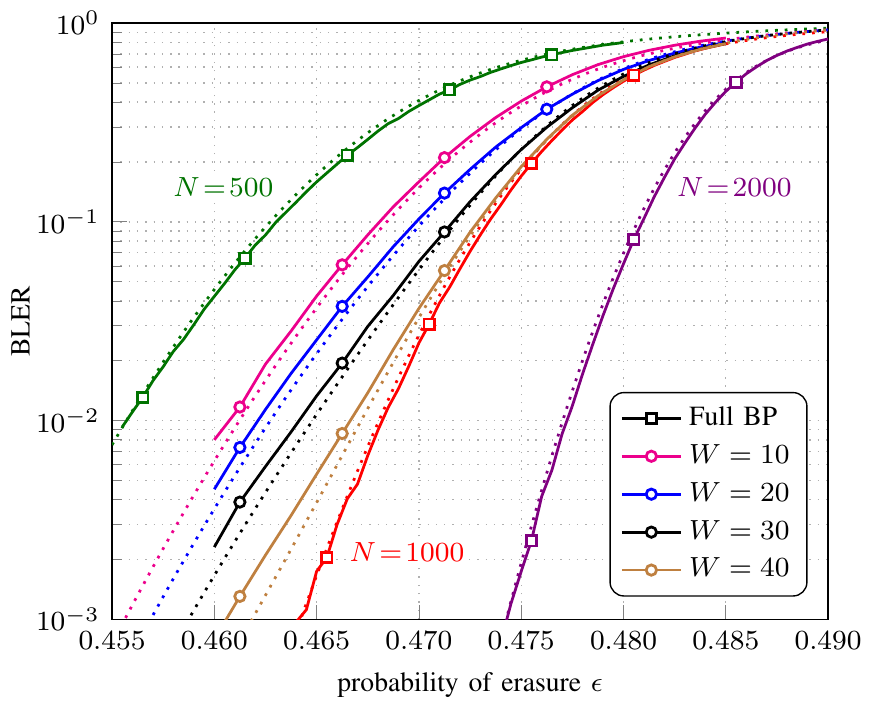}
    \vspace{-15pt}
    \caption{Simulated BLER curves (solid lines with circles) and corresponding analytical approximations (dotted
    lines) for the terminated $(5,10,L\!=\!50,N\!=\!1000)$ ensemble for different values of the window size $W$.
    The simulated BLER performance of the $(5,10,L\!=\!50,N)$ ensemble under full BP decoding for different values of $N$ is also shown (solid lines with squares) alongside the corresponding analytical approximations (dotted lines).}
    \label{fig:BLER_5_10}
    \vspace{-2ex}
\end{figure}

\tikzsetnextfilename{bec_5_10_sw_ber_latency}
\begin{figure}[!t]
    \centering
    \includegraphics[width=\columnwidth]{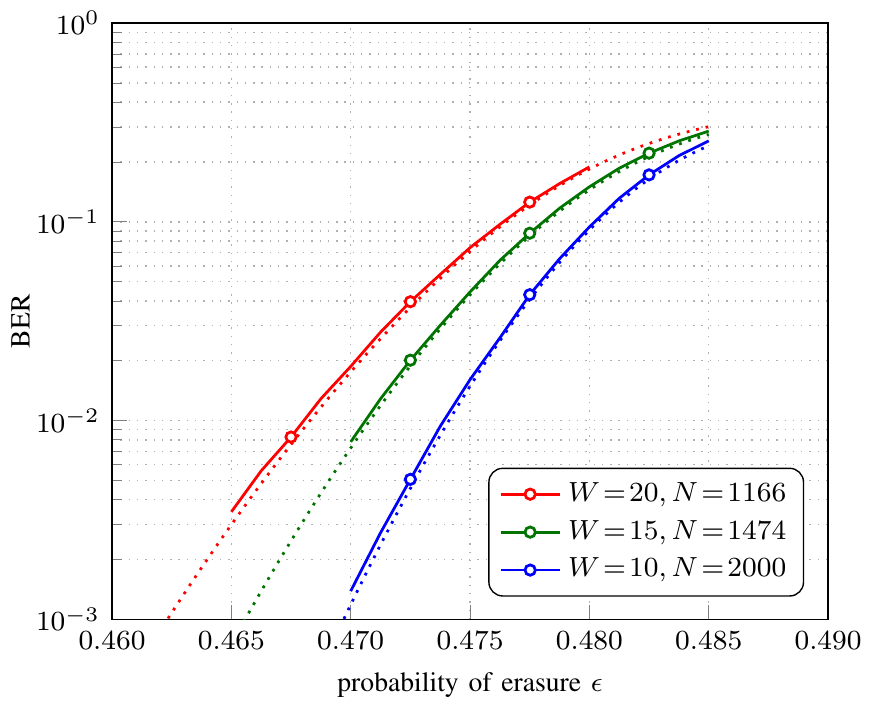}
    \vspace{-15pt}
    \caption{Simulated BER curves (solid lines) and corresponding analytical approximations (dotted
    lines) for the terminated $(5,10,L\!=\!50,N)$ ensemble under sliding window decoding. The values of $W$ and $N$ are chosen to keep the decoding latency approximately equal to \mbox{$28\!\cdot\!10^3$} bits.}
    \label{fig:sw_latency}
    \vspace{-2ex}
\end{figure}

In Fig.~\ref{fig:our_approx_5_10} we compare the simulated FER and BER performance with the
analytical approximations in~\eqref{eq:our_fer} and~\eqref{eq:our_ber} for the terminated
${(5,10,L\!=\!50,N)}$ SC-LDPC code ensemble with $N=500$, $1000$, and $2000$.
The refined scaling law predicts the frame and
bit error rates very accurately.
The prediction of the FER via the scaling law in~\cite{Olmo15}
(see Section~\ref{sse:old}) is also shown for comparison.
It shows a significant gap to the simulated curves, which is closed by the proposed refined
scaling law.
We remark that since we are only interested in large (linear sized with respect to $N$) error
events, we ignore all failures involving only size-$2$ stopping sets when calculating the simulated
error rates, effectively considering an expurgated ensemble, similar to the approach used
in~\cite{Amra09} to remove the effect of the error floor.

Introducing the dependence of $\sstart$, $\ssend$, and $\gamma$ on $\epsilon,$ which we discussed in Section~\ref{sse:additional_refinements}, slightly improves the prediction of the FER.
Fig.~\ref{fig:refinements} shows this improvement for the terminated \mbox{$(5,10,L\!=\!50,N\!=\!2000)$} SC-LDPC code ensemble.
The black curve with squares shows the prediction made using~\eqref{eq:our_fer} with parameters $(\sstartlb = 0.0053 L, \ssenddouble = \e L, \gammasingle = \gammadouble / 2 = 4.19 / 2,\nusingle=0.424,\thetasingle=1.64)$, i.e., with $\sstart$ and $\gamma$ modeled as $\e$-independent constants and $\ssend$ upper bounded by $\e L$ (the coefficients $\sstartlb / L$ and $\gammadouble$ are taken from~\cite{Olmo15}).
The prediction that models $(\sstartdouble,\ssenddouble,\gammasingle)$ as dependent on $\e$ (red dashed curve) is in a better agreement with the simulated FER (blue curve).

On the other hand, the prediction of the original scaling law in~\cite{Olmo15} does not improve when parameters $\sstart$, $\ssend$, and $\gamma$ are modeled as functions of $\e$.
Indeed, the black line with diamonds in Fig.~\ref{fig:refinements} is obtained using~\eqref{eq:olmos_fer} as in~\cite{Olmo15} with scaling parameters $(\sstartlb\!=\!0.0053 L,\ssenddouble\!=\!\e L,\gammadouble\!=\!4.19,\nudouble\!=\!2\!\cdot\!\nusingle=\!2\!\cdot\!0.424,\thetadouble\!=\!0.63)$.
The green dash-dotted line corresponds to the prediction~\eqref{eq:olmos_fer} with the same $\nudouble$ and $\thetadouble$ (the value of the correlation decay constant $\thetadouble$ is taken from~\cite{Olmo15}) but with $\e$-dependent $\sstartdouble,\ssenddouble$, and $\gammadouble$.
We observe that introducing the dependence worsens the prediction by the scaling law in~\cite{Olmo15}.
This indicates that the mismatch between the prediction in~\cite{Olmo15} and simulation results is due to modeling the decoding process by a single Ornstein-Uhlenbeck process, instead of two processes as proposed here.

The simulated and predicted FER and BER performance of the terminated ${(4,8,L\!=\!50,N)}$ SC-LDPC
code ensemble is shown in Fig.~\ref{fig:our_approx_4_8_term}.
Similar to the case of the ${(5,10,L\!=\!50,N)}$ ensemble, a very good match is observed
between the curves predicted by the refined scaling law and the simulation results.

The results for the terminated $(3,6,L\!=\!50,N)$ SC-LDPC code ensemble are given
in Fig.~\ref{fig:our_approx_3_6_term}.
We observe that in this case the performance curves predicted by the refined scaling law are
characterized by a gap to the corresponding simulation results, although the gap is considerably
smaller than for the scaling law in~\cite{Olmo15}.
We remark that the predicted error rates are very accurate for the ensembles with VN degree $\dv\!\ge\!4$
irrespective of the code rate,
whereas for $\dv\!=\!3$ a gap appears, albeit significantly smaller than that for the original
scaling law in~\cite{Olmo15}.

In Figs.~\ref{fig:FER_5_10},~\ref{fig:BER_5_10}, and~\ref{fig:BLER_5_10} we compare the simulated frame, bit, and block error rates with the analytical approximations for the terminated
${(5,10,L=50,N=1000)}$ SC-LDPC code ensemble under sliding window decoding and window sizes $W=10,20,30$, and $40$.
A very good agreement between the analytical results and the simulation results is observed for all window sizes.
We remark that for a given $W$ the accuracy of the prediction increases with $L$.
Besides the BLER curves for the sliding window decoding, Fig.~\ref{fig:BLER_5_10} shows that the prediction of the BLER for full BP decoding in~\eqref{eq:our_bler} is as accurate as that of the FER and BER in Fig.~\ref{fig:our_approx_5_10}.

Finally, Fig.~\ref{fig:sw_latency} illustrates how to use the scaling law in~\eqref{eq:PbSWb} to jointly design the size of the sliding window $W$ and the component code length $N$.
We consider the terminated $(5,10,L\!=\!50,N)$ SC-LDPC code ensemble under sliding window decoding and set the decoding latency $N(W+\dv-1)$ to approximately  $28\!\cdot\!10^3$ bits.
Fig.~\ref{fig:sw_latency} shows the BER curves for three combinations of~$W$ and $N$ that (approximately) yield this latency.
The curves indicate that once the sliding window is made large enough to accommodate the decoding wave, the code designer should opt for a larger $N$ to increase the slope of the error rate curve.
We observe again that the scaling law accurately predicts the simulated BER curves, allowing the code designer to choose $W$ and $N$ without having to resort to Monte-Carlo simulations for each combination.

\section{Conclusion}
\label{sec:conclusion}

We proposed a finite-length scaling law for SC-LDPC codes decoded using window decoding over the binary erasure channel. 
We first proposed a scaling law for terminated SC-LDPC codes under full belief propagation decoding by modeling the decoding process as two independent Ornstein-Uhlenbeck processes,
corresponding to the two decoding waves moving from the boundaries toward the center of the coupled
chain. This scaling law is a refinement of the law proposed by Olmos and Urbanke, which considers a single Ornstein-Uhlenbeck process, and yields a much more accurate prediction of the error rate. We then extended the proposed scaling law to the more interesting case of window decoding by modeling the decoding as a two-phase process, the first phase characterized by a single decoding wave and the second by two decoding waves.
The proposed scaling law provides a very accurate prediction of the frame, bit, and block error rate
performance for SC-LDPC codes of variable node degree larger than or equal to~$4$ under full BP
decoding and under window decoding with window size at least $10$.
For variable node degree $3$, a small gap remains.
Closing it is an interesting research problem, which may be partially addressed by considering the possibility of a decoding failure outside of the steady-state region.

Using the scaling law, we can easily estimate the price to be paid in terms of error-correcting performance when using window
decoding instead of full BP.
Notably, as Fig.~\ref{fig:FER_5_10} shows, this price is not negligible even for large windows---the slopes of
the error rate curves differ.
For window decoding, irrespective of the size of the window, the error probability is dominated by the single-wave
phase, making it impossible to achieve the performance of full two-wave BP decoding.



\begin{thebibliography}{10}
\providecommand{\url}[1]{#1}
\csname url@samestyle\endcsname
\providecommand{\newblock}{\relax}
\providecommand{\bibinfo}[2]{#2}
\providecommand{\BIBentrySTDinterwordspacing}{\spaceskip=0pt\relax}
\providecommand{\BIBentryALTinterwordstretchfactor}{4}
\providecommand{\BIBentryALTinterwordspacing}{\spaceskip=\fontdimen2\font plus
\BIBentryALTinterwordstretchfactor\fontdimen3\font minus
  \fontdimen4\font\relax}
\providecommand{\BIBforeignlanguage}[2]{{%
\expandafter\ifx\csname l@#1\endcsname\relax
\typeout{** WARNING: IEEEtran.bst: No hyphenation pattern has been}%
\typeout{** loaded for the language `#1'. Using the pattern for}%
\typeout{** the default language instead.}%
\else
\language=\csname l@#1\endcsname
\fi
#2}}
\providecommand{\BIBdecl}{\relax}
\BIBdecl

\bibitem{Jime99}
A.~{Jimen\'ez Feltstr\"om} and K.~S. {Zigangirov}, ``Time-varying periodic
  convolutional codes with low-density parity-check matrix,'' \emph{IEEE Trans.
  Inf. Theory}, vol.~45, no.~6, pp. 2181--2191, Sep. 1999.

\bibitem{Lent10}
M.~{Lentmaier}, A.~{Sridharan}, D.~J. {Costello}, and K.~S. {Zigangirov},
  ``Iterative decoding threshold analysis for {LDPC} convolutional codes,''
  \emph{IEEE Trans. Inf. Theory}, vol.~56, no.~10, pp. 5274--5289, Oct. 2010.

\bibitem{Kude11}
S.~Kudekar, T.~J. Richardson, and R.~L. Urbanke, ``Threshold saturation via
  spatial coupling: Why convolutional {LDPC} ensembles perform so well over the
  {BEC},'' \emph{IEEE Trans. Inf. Theory}, vol.~57, no.~2, pp. 803--834, Feb.
  2011.

\bibitem{Kude13}
S.~{Kudekar}, T.~{Richardson}, and R.~L. {Urbanke}, ``Spatially coupled
  ensembles universally achieve capacity under belief propagation,'' \emph{IEEE
  Trans. Inf. Theory}, vol.~59, no.~12, pp. 7761--7813, Dec. 2013.

\bibitem{Srid07}
A.~{Sridharan}, D.~{Truhachev}, M.~{Lentmaier}, D.~J. {Costello}, and K.~S.
  {Zigangirov}, ``Distance bounds for an ensemble of {LDPC} convolutional
  codes,'' \emph{IEEE Trans. Inf. Theory}, vol.~53, no.~12, pp. 4537--4555,
  Dec. 2007.

\bibitem{Cost14}
D.~J. {Costello}, L.~{Dolecek}, T.~E. {Fuja}, J.~{Kliewer}, D.~G.~M.
  {Mitchell}, and R.~{Smarandache}, ``Spatially coupled sparse codes on graphs:
  theory and practice,'' \emph{IEEE Commun. Mag.}, vol.~52, no.~7, pp.
  168--176, Jul. 2014.

\bibitem{Molo17}
S.~{Moloudi}, M.~{Lentmaier}, and A.~{Graell i Amat}, ``Spatially coupled
  turbo-like codes,'' \emph{IEEE Trans. Inf. Theory}, vol.~63, no.~10, pp.
  6199--6215, Oct. 2017.

\bibitem{Smit12}
B.~P. {Smith}, A.~{Farhood}, A.~{Hunt}, F.~R. {Kschischang}, and J.~{Lodge},
  ``Staircase codes: {FEC} for 100 {Gb}/s {OTN},'' \emph{J. Lightw. Technol.},
  vol.~30, no.~1, pp. 110--117, Jan. 2012.

\bibitem{Are12}
V.~Aref, N.~Macris, R.~Urbanke, and M.~Vuffray, ``Lossy source coding via
  spatially coupled {LDGM} ensembles,'' in \emph{Proc. IEEE Int. Symp. Inf.
  Theory (ISIT)}, Cambridge, MA, Jul. 2012, pp. 373 --377.

\bibitem{Don13}
D.~L. Donoho, A.~Javanmard, and A.~Montanari, ``Information-theoretically
  optimal compressed sensing via spatial coupling and approximate message
  passing,'' \emph{IEEE Trans. Inf. Theory}, vol.~59, no.~11, pp. 7434--7464,
  Nov. 2013.

\bibitem{Iye12}
A.~R. {Iyengar}, M.~{Papaleo}, P.~H. {Siegel}, J.~K. {Wolf},
  A.~{Vanelli-Coralli}, and G.~E. {Corazza}, ``Windowed decoding of
  protograph-based {LDPC} convolutional codes over erasure channels,''
  \emph{IEEE Trans. Inf. Theory}, vol.~58, no.~4, pp. 2303--2320, Apr. 2012.

\bibitem{Amra09}
A.~Amraoui, A.~Montanari, T.~Richardson, and R.~Urbanke, ``Finite-length
  scaling for iteratively decoded {LDPC} ensembles,'' \emph{IEEE Trans. Inf.
  Theory}, vol.~55, no.~2, pp. 473--498, Feb. 2009.

\bibitem{Ezri08_slope}
J.~{Ezri}, A.~{Montanari}, S.~{Oh}, and R.~{Urbanke}, ``The slope scaling
  parameter for general channels, decoders, and ensembles,'' in \emph{Proc.
  IEEE Int. Symp. Inf. Theory (ISIT)}, Toronto, Canada, Jul. 2008, pp.
  1443--1447.

\bibitem{Ezri08_shift}
J.~{Ezri}, R.~{Urbanke}, A.~{Montanari}, and {Sewoong Oh}, ``Computing the
  threshold shift for general channels,'' in \emph{Proc. IEEE Int. Symp. Inf.
  Theory (ISIT)}, Toronto, Canada, Jul. 2008, pp. 1448--1452.

\bibitem{Olmo15}
P.~M. Olmos and R.~L. Urbanke, ``A scaling law to predict the finite-length
  performance of spatially-coupled {LDPC} codes,'' \emph{IEEE Trans. Inf.
  Theory}, vol.~61, no.~6, pp. 3164--3184, Jun. 2015.

\bibitem{Stin16}
M.~Stinner and P.~M. Olmos, ``On the waterfall performance of finite-length
  {SC}-{LDPC} codes constructed from protographs,'' \emph{IEEE J. Sel. Areas
  Commun.}, vol.~34, no.~2, pp. 345--361, Feb. 2016.

\bibitem{Cost18}
D.~J. {Costello}, D.~G.~M. {Mitchell}, P.~M. {Olmos}, and M.~{Lentmaier},
  ``Spatially coupled generalized {LDPC} codes: Introduction and overview,'' in
  \emph{Proc. 10th IEEE Int. Symp. Turbo Codes and Iterative Inf. Process.
  (ISTC)}, Hong Kong, China, Dec. 2018.

\bibitem{Luby97}
M.~G. Luby, M.~Mitzenmacher, M.~A. Shokrollahi, D.~A. Spielman, and V.~Stemann,
  ``Practical loss-resilient codes,'' in \emph{Proc. 29th Annu. ACM Symp.
  Theory Comput.}, El Paso, TX, USA, May 1997, pp. 150--159.

\bibitem{Luby01}
M.~G. {Luby}, M.~{Mitzenmacher}, M.~A. {Shokrollahi}, and D.~A. {Spielman},
  ``Efficient erasure correcting codes,'' \emph{IEEE Trans. Inf. Theory},
  vol.~47, no.~2, pp. 569--584, Feb. 2001.

\end{thebibliography}
\end{document}